# Comment on "Transformation of phase velocity among inertial frames" [Wave Motion 50 (2013) 520-524]

Seyed Saied Mirahmadi*
*Corresponding author: mirrahmadi@bou.ac.ir
ORCID: https://orcid.org/0000-0002-7942-0109

**Abstract**
The transformation formulas between two inertial frames for the minimum phase velocity and the phase velocity in an arbitrary direction are given. The derived transformation formula for the reciprocal of the phase velocity along the particle's direction of motion is a correction of the formula provided by J.J. Zhang and P.X. Wang in Wave Motion 50 (2013) 520-524.

The velocity of the phase of a wave traveling in space is called the phase velocity which is a significant concept for describing the wave propagation and a useful quantity in applications [1-3]. The phase velocity [4] and the minimum phase velocity [4-6] transformation formulas are important in some applications such as interactions between waves and moving particles and signal communications between moving sources and Earth-based receivers.

On any cophasal surface, the phase $\varphi(\boldsymbol{r},t)$ remains unchanged, That is, $d\varphi = \boldsymbol{\nabla}\varphi \,.\, d\boldsymbol{r} + (\partial\varphi/\partial t)dt = 0$. The phase velocity thus satisfies the following equation

$$-\boldsymbol{\nabla}\varphi.\boldsymbol{v}_p = \partial\varphi/\partial t. \tag{1}$$

From this equation, we obtain the phase velocity in an arbitrary direction $\boldsymbol{j}$:

$$\boldsymbol{v}_p = \frac{\omega\,\boldsymbol{j}}{\boldsymbol{k}\,.\,\boldsymbol{j}} \qquad or \qquad \boldsymbol{v}_p = \frac{\omega\,\hat{\boldsymbol{j}}}{-\boldsymbol{\nabla}_{\hat{j}}\varphi}\,, \tag{2}$$

where $\omega = \partial\varphi/\partial t$ and $\boldsymbol{k} = -\boldsymbol{\nabla}\varphi$ are the angular frequency and the angular wave number vector, respectively and $\boldsymbol{\nabla}_{\hat{j}}\varphi = \boldsymbol{\nabla}\varphi.\hat{\boldsymbol{j}}$ is the gradient of the phase field in the direction of $\hat{\boldsymbol{j}}$ which is the unit vector of $\boldsymbol{j}$.

The minimum phase velocity occurs in the direction of $\boldsymbol{k}$, since the maximum value for $\boldsymbol{k}\,.\,\boldsymbol{j}$ occurs,

$$\boldsymbol{v}_{p\,min} = \frac{\omega\,\boldsymbol{k}}{\boldsymbol{k}\,.\,\boldsymbol{k}}\,, \tag{3}$$

Let us consider that the frame $S'$ is moving, without any rotation, with a velocity $\boldsymbol{v}$ with respect to an inertial frame $S$ and further assume that the corresponding axes of the frames are parallel to each other. For the same argument presented above, the mathematical form of Eqs. (1)-(3) are applicable for the phase field in the moving frame $\varphi'(\boldsymbol{r}',t')$, or in other words these equations are covariant (form-invariant) under Lorentz transformation. Therefore, we can obtain the phase velocity in an arbitrary direction $\boldsymbol{j}'$ and the minimum phase velocity in the moving frame $S'$ respectively as

$$\boldsymbol{v}'_p = \frac{\omega'\boldsymbol{j}'}{\boldsymbol{k}'.\boldsymbol{j}'} \qquad or \qquad \boldsymbol{v}'_p = \frac{\omega'\,\widehat{\boldsymbol{j}'}}{-\boldsymbol{\nabla}'_{\widehat{j'}}\varphi'}\,, \tag{4}$$

and

$$\boldsymbol{v}'_{p\,min} = \frac{\omega'\boldsymbol{k}'}{\boldsymbol{k}'.\boldsymbol{k}'}\,, \tag{5}$$

where $\omega' = \partial\varphi'/\partial t'$ and $\boldsymbol{k}' = -\boldsymbol{\nabla}'\varphi'$ are the angular frequency and the angular wave number vector in the frame $S'$, respectively and $\boldsymbol{\nabla}'_{\hat{j}'}\varphi' = \boldsymbol{\nabla}'\varphi'.\,\hat{\boldsymbol{j}}'$ is the gradient of the phase field in the direction of $\hat{\boldsymbol{j}}'$ which is the unit vector of $\boldsymbol{j}'$.

Although the derivation of the phase field in the moving frame $\varphi'(\boldsymbol{r}', t')$ to calculate the $\omega' = \frac{\partial\varphi'}{\partial t'}$ and $\boldsymbol{k}' = -\boldsymbol{\nabla}'\varphi'$ is not easy in most cases, we can obtain $\omega'$ and $\boldsymbol{k}'$ by applying the phase invariance and the Lorentz covariance of the phase. So, for a monochromatic plane wave

$$\varphi(\boldsymbol{r}, t) = \omega\, t - \boldsymbol{k}.\boldsymbol{r} = \omega'\, t' - \boldsymbol{k}'.\boldsymbol{r}' = \varphi'(\boldsymbol{r}', t'). \tag{6}$$

Rewriting Eq. (6) in 4-tensor form

$$K_\mu\, x^\mu = K_{\alpha'}\, x^{\alpha'}, \tag{7}$$

where $x^\mu$ , $x^{\alpha'}$ , $K_\mu$ and $K_{\alpha'}$ are the components of the position 4-vectors $\mathbf{X} = (ct, x, y, z)$ and $\mathbf{X}' = (ct', x', y', z')$, and the components of the wave number 4-vectors $\mathbf{K} = \left(-\frac{\omega}{c}, \boldsymbol{k}\right)$ and $\mathbf{K}' = \left(-\frac{\omega'}{c}, \boldsymbol{k}'\right)$, respectively, where $c$ is the speed of light in vacuum, and applying Lorentz transformation

$$x^{\alpha'} = \mathrm{L}^{\alpha'}{}_{\mu}\, x^\mu, \tag{8}$$

where $\mathrm{L}^{0'}{}_{0} = \gamma$, $\mathrm{L}^{0'}{}_{i} = \mathrm{L}^{i'}{}_{0} = -\gamma\frac{v^i}{c}$, $\mathrm{L}^{i'}{}_{j} = \mathrm{L}^{j'}{}_{i} = \delta^{ij} + (\gamma - 1)\frac{v^i v^j}{v^2}$ $(i, j = 1 - 3)$ and $\gamma = \frac{1}{\sqrt{1 - v^2/c^2}}$, to Eq. (7), we get, after utilizing $\mathrm{L}^{\alpha'}{}_{\mu}\,\mathrm{L}^{\mu}{}_{\beta'} = \delta^{\alpha'}{}_{\beta'}$,

$$K_{\alpha'} = K_\mu \mathrm{L}^{\mu}{}_{\alpha'}\,, \tag{9}$$

where $\mathrm{L}^{0}{}_{0'} = \mathrm{L}^{0'}{}_{0}$, $\mathrm{L}^{i}{}_{0'} = \mathrm{L}^{0}{}_{i'} = -\mathrm{L}^{0'}{}_{i} = -\mathrm{L}^{i'}{}_{0}$ and $\mathrm{L}^{i}{}_{j'} = \mathrm{L}^{j}{}_{i'} = \mathrm{L}^{i'}{}_{j} = \mathrm{L}^{j'}{}_{i}$.

Eq. (9) can be written in 3-vector form as

$$\omega' = \gamma\omega - \gamma\boldsymbol{k}\,.\boldsymbol{v}, \tag{10}$$

$$\boldsymbol{k}' = \boldsymbol{k} + \left[(\gamma - 1)\frac{\boldsymbol{k}.\boldsymbol{v}}{v^2} - \gamma\frac{\omega}{c^2}\right]\boldsymbol{v}. \tag{11}$$

Substituting from Eq. (10), Eq. (11) and $\boldsymbol{k} = \frac{\omega\, \boldsymbol{v}_{p\,min}}{(v_{p\,min})^2}$, which follows directly from Eq. (3), into Eq. (5) and after some manipulations, we find the minimum phase velocity transformation formula as

$$\boldsymbol{v}'_{p\,min} = \frac{\gamma\left(1 - \frac{\boldsymbol{v}_{p\,min}\,.\boldsymbol{v}}{(v_{p\,min})^2}\right)\left[\boldsymbol{v}_{p\,min} + \left((\gamma - 1)\frac{\boldsymbol{v}_{p\,min}.\boldsymbol{v}}{v^2} - \gamma\frac{(v_{p\,min})^2}{c^2}\right)\boldsymbol{v}\right]}{1 - \frac{(v_{p\,min})^2}{c^2}\left[1 - \gamma^2\left(1 - \frac{\boldsymbol{v}_{p\,min}\,.\boldsymbol{v}}{(v_{p\,min})^2}\right)^2\right]}. \tag{12}$$

According to Eq. (4), in order to derive the transformation formula for $\boldsymbol{v}'_p$, in addition to the $\boldsymbol{k}'$ transformation formula, Eq. (11), we need to know the transformation formula for $\boldsymbol{j}'$. For instance, we can obtain the phase velocity in the moving frame $S'$ in the direction of the particle's velocity ($\boldsymbol{j}' = \boldsymbol{u}'$) by applying Eqs. (4), (10) and (11),

$$\boldsymbol{v}'_p = \frac{\omega'\boldsymbol{u}'}{\boldsymbol{k}'.\boldsymbol{u}'}, \tag{13}$$

where

$$\boldsymbol{u}' = \frac{1}{1 - \frac{\boldsymbol{u}\,.\boldsymbol{v}}{c^2}}\left[\frac{1}{\gamma}\boldsymbol{u} - \left(1 - \frac{\gamma}{\gamma + 1}\frac{(\boldsymbol{u}\,.\boldsymbol{v})}{c^2}\right)\boldsymbol{v}\right]. \tag{14}$$

One can find $\boldsymbol{k}'.\boldsymbol{u}'$ in Eq. (13) by applying Eqs. (11) and (14). However, another way for calculating $\boldsymbol{k}'.\boldsymbol{u}'$ is applying 4-velocities $\mathbf{U} = \gamma_u(c, u_x, u_y, u_z)$, where $\gamma_u = \frac{1}{\sqrt{1-u^2/c^2}}$, and $\mathbf{U}' = \gamma_{u'}(c, u'_x, u'_y, u'_z)$, where $\gamma_{u'} = \frac{1}{\sqrt{1-u'^2/c^2}} = \gamma\gamma_u(1 - \boldsymbol{u}\,.\,\boldsymbol{v}/c^2)$, in frames $S$ and $S'$ respectively and using the 4-velocities transformation formula

$$U^{\alpha'} = \mathrm{L}^{\alpha'}{}_{\mu} U^{\mu}. \tag{15}$$

From Eqs. (15) and (9) we have

$$U^{\alpha'} K_{\alpha'} = \mathrm{L}^{\alpha'}{}_{\mu} U^{\mu} K_{\beta} \mathrm{L}^{\beta}{}_{\alpha'} = U^{\mu} K_{\mu}\,. \tag{16}$$

Substituting from Eq. (16) into the following equation

$$\boldsymbol{k}'.\boldsymbol{u}' = \frac{1}{\gamma_{u'}}\left(U^{\alpha'} K_{\alpha'} - U^{0'} K_{0'}\right) = \frac{1}{\gamma_{u'}}\left(U^{\mu} K_{\mu} - \mathrm{L}^{0'}{}_{\mu} U^{\mu} K_{\beta} \mathrm{L}^{\beta}{}_{0'}\right), \tag{17}$$

after a little manipulation, yields

$$\boldsymbol{k}'.\boldsymbol{u}' = \frac{\omega}{\gamma\left(1 - \frac{\boldsymbol{u}\,.\,\boldsymbol{v}}{c^2}\right)}\left(\frac{u}{v_p} - 1\right) + \omega', \tag{18}$$

where $v_p$ is the phase velocity value in the direction of the particle's velocity in frame $S$. Substituting Eq. (18) into Eq. (13), we obtain the phase velocity in the moving frame in the direction of the particle's velocity ($\boldsymbol{j}' = \boldsymbol{u}'$) as

$$\boldsymbol{v}'_p = \frac{\omega' \boldsymbol{u}'}{\frac{\omega}{\gamma\left(1 - \frac{\boldsymbol{u}\,.\,\boldsymbol{v}}{c^2}\right)}\left(\frac{u}{v_p} - 1\right) + \omega'} = \frac{\omega' |\boldsymbol{u}'|}{\frac{\omega}{\gamma\left(1 - \frac{\boldsymbol{u}\,.\,\boldsymbol{v}}{c^2}\right)}\left(\frac{u}{v_p} - 1\right) + \omega'} \frac{\boldsymbol{u}'}{|\boldsymbol{u}'|}, \tag{19}$$

or, by substituting $|\boldsymbol{u}'|$, which follows directly from Eq. (14), into Eq. (19),

$$\boldsymbol{v}'_p = \frac{c\sqrt{\gamma^2\gamma_u^2\left(1 - \frac{\boldsymbol{u}\,.\,\boldsymbol{v}}{c^2}\right)^2 - 1}}{\frac{\omega}{\omega'}\gamma_u\left(\frac{u}{v_p} - 1\right) + \gamma\gamma_u\left(1 - \frac{\boldsymbol{u}\,.\,\boldsymbol{v}}{c^2}\right)} \frac{\boldsymbol{u}'}{|\boldsymbol{u}'|}, \tag{20}$$

where $\boldsymbol{u}'$ and $\omega'$ are given by Eqs. (14) and (10) respectively.

From Eq. (19) we get the reciprocal of the phase velocity in the moving frame $S'$ along the particle's direction of motion as

$$\Lambda'_p = \frac{1}{v'_p} = \frac{\frac{\omega}{\gamma\left(1 - \frac{\boldsymbol{u}\,.\,\boldsymbol{v}}{c^2}\right)}\left(u\,\Lambda_p - 1\right) + \omega'}{\omega' |\boldsymbol{u}'|}, \tag{21}$$

and from Eq. (20)

$$\Lambda'_p = \frac{1}{c}\frac{\frac{\omega}{\omega'}\gamma_u\left(u\,\Lambda_p - 1\right) + \gamma\gamma_u\left(1 - \frac{\boldsymbol{u}\,.\,\boldsymbol{v}}{c^2}\right)}{\sqrt{\gamma^2\gamma_u^2\left(1 - \frac{\boldsymbol{u}\,.\,\boldsymbol{v}}{c^2}\right)^2 - 1}}, \tag{22}$$

where $\Lambda_p = 1/v_p$.

According to Eqs. (21) and (22), it is obvious that Eq. (17) in Zhang and Wang's paper [4] is incorrect. Furthermore, the dimensional analysis reveals a defect in their Eq. (17) and just above this equation, their statement is not correct. Utilizing their Eqs. (7), (8) and $\boldsymbol{j}' = \frac{\boldsymbol{u}'}{|\boldsymbol{u}'|}$, one gets

$$\Lambda'_p = \frac{\boldsymbol{k}'.\boldsymbol{u}'}{\omega' |\boldsymbol{u}'|}. \tag{23}$$

So, according to Eq. (23) and their Eq. (16), the following expression is the correction of their statement: "In frame $S'$, the reciprocal of the phase velocity along the particle's direction of motion is the spatial part of the above-mentioned projection divided by $\gamma_u \omega' |\boldsymbol{u}'|$."

Utilizing Eq. (10) of Ref [4], one obtains $\omega' = \gamma\omega - \gamma\boldsymbol{k}\,.\,\boldsymbol{v}$ which shows that the term "$\gamma\boldsymbol{k}\,.\,\boldsymbol{v}$" in Eq. (19) in Ref. [4] should be corrected as "$-\gamma\boldsymbol{k}\,.\,\boldsymbol{v}$".

In the paragraph following Eq. (19) of Ref. [4], the authors concluded: "However, in frame $S'$, $\boldsymbol{v}$ becomes a zero vector, and the projection of the wave vector in this direction becomes zero; namely, the reciprocal of the phase velocity along this zero vector is also zero. Thus, in frame $S'$, the phase velocity along the direction $\boldsymbol{j} = \frac{1}{v}(v_x, v_y, v_z)$ in frame $S$ becomes infinite." This conclusion is not correct. In frame $S'$, $\boldsymbol{v}$ becomes a zero vector and it is obvious that the reciprocal of the phase velocity along the direction of a *zero* vector (i.e. $\boldsymbol{v}$) is meaningless not zero. Thus, it is *not* the case that in frame $S'$ the phase velocity along the direction $\boldsymbol{j} = \frac{1}{v}(v_x, v_y, v_z)$ in frame $S$ becomes necessarily infinite.

## Statements and Declarations

*Funding:* The author declares that no funds, grants, or other support were received during the preparation of this manuscript.

*Competing Interests:* The author has no relevant financial or non-financial interests to disclose.

*Availability of data and materials:* Data presented in the manuscript are available via request.